\documentclass[11pt, a4paper]{article}
\usepackage{amsmath}

\newtheorem{theorem}{Theorem}

\newcommand{\benumerate}{\begin{enumerate}}
\newcommand{\eenumerate}{\end{enumerate}}

\newcommand{\bitemize}{\begin{itemize}}

\newcommand{\eitemize}{\end{itemize}}
\newcommand{\ep}{\epsilon}

\begin{document}

\title{Towards the classification of  integrable differential-difference equations in $2+1$ dimensions }
\author{E.V. Ferapontov, V.S. Novikov, I. Roustemoglou}
    \date{}
    \maketitle
    \vspace{-7mm}
\begin{center}
Department of Mathematical Sciences \\ Loughborough University \\
Loughborough, Leicestershire LE11 3TU \\ United Kingdom \\[2ex]
e-mails: \\
 { \texttt{E.V.Ferapontov@lboro.ac.uk}}\\
  {\texttt{V.Novikov@lboro.ac.uk}}\\
  { \texttt{I.Roustemoglou@lboro.ac.uk}}\\
\end{center}

\bigskip

\begin{abstract}

We address the problem of classification of integrable differential-difference equations in 2+1 dimensions with one/two discrete variables. Our approach is based on the method of hydrodynamic reductions and its generalisation to dispersive equations as proposed in  \cite{FerM, FMN}. We obtain a number of classification results of scalar integrable equations including that of  the intermediate long wave and Toda type.

\bigskip


\bigskip

\end{abstract}

\newpage

\section{Introduction}

This paper is aimed at  the classification of scalar $2+1$ dimensional integrable equations of the general form
\begin{equation}
u_t=F(u, w),
\label{1}
\end{equation}
where $u(x, y, t)$ is a scalar field, $w(x, y, t)$ is the nonlocal variable,  and  $F$ is a differential/difference operator in the independent variables $x$ and $y$ (the explicit form of $w$ and $F$ will be specified in what follows). Publications \cite{FerM, FMN} provide a novel perturbative approach to the classification of integrable  equations  of the form (\ref{1}) which have  non-degenerate dispersionless limit. This approach is based on the requirement that all hydrodynamic reductions \cite{Fer4} of the dispersionless limit can be deformed into reductions of the full dispersive equation (a brief description of this method is included in Sect. 2). For the simplest nonlocality $w=D_x^{-1}D_yu$ (or, equivalently, $w_x=u_y$), the paper \cite{FMN} gives a complete list of integrable third order PDEs of the form
\begin{equation}
\label{nonsym}
\begin{aligned}
u_t&=\varphi u_x+\psi u_y +\eta w_y+\epsilon(...)+\epsilon^2(...),
\end{aligned}
\end{equation}
where  $\varphi, \psi$ and $ \eta$ are functions of $u$ and $w$, while
the terms at  $\epsilon$ and $\epsilon^2$ are assumed to be
homogeneous differential polynomials of  the order two and  three
in the $x-$ and $y-$derivatives of $u$ and $w$.

\begin{theorem}\cite{FMN} Up to invertible transformations, the examples below provide a  complete   list of integrable third order equations (\ref{nonsym}) with non-degenerate dispersionless limit:
\begin{align*}
 &{ KP ~ equation}& \qquad &
 u_{t} = u u_{x} +w_y{{+ \epsilon^{2} u_{xxx}}}, &\\
&{ modified ~ KP ~ equation}& \qquad &
u_{t} =(w- \frac{u^2}{2})  u_{x} +w_y {{+\epsilon^2  u_{xxx}}},&\\
&{Gardner ~ equation}& \qquad &
 u_{t} =(\beta w- \frac{\beta^2}{2}u^2+\delta u)  u_{x} +w_y {{+\epsilon^2  u_{xxx}}},&\\
&{  VN ~ equation}& \qquad &
u_{t} =(u w)_{y} {{+\epsilon^2  u_{yyy}}},&\\
&{ modified ~ VN ~ equation}& \qquad &
u_{t} =(u w)_y {{+\epsilon^2  \left( u_{yy}-\frac{3}{4}\frac{u_y^2}{u}\right)_y}},&\\
&{HD ~ equation }& \qquad &
u_{t} =-2w  u_{y} +u w_y {{-\frac{\epsilon^2}{u}\left(\frac{1}{u}\right)_{xxx}}},\\
&{ deformed ~ HD~ equation }& \qquad &
u_{t} =\frac{\delta}{u^3}u_x-2w  u_{y} +u w_y {{-\frac{\epsilon^2}{u}\left(\frac{1}{u}\right)_{xxx}}},&\\
&{ Equation \ E_1 }& \qquad &
u_t=(\beta w+\beta^2u^2)u_x-3\beta u u_y+w_y+\epsilon^2
[B^3(u)-\beta u_x B^2(u)],&\\
&{Equation \ E_2 }& \qquad &
u_t=\frac{4}{3}\beta^2 u^3u_x+(w-3\beta u^2) u_y+uw_y+\ep^2
[B^3(u)-\beta u_xB^2(u)],&
\end{align*}
here $B=\beta u D_x-D_y$, $\beta, \delta$=const.
\end{theorem}

\noindent Although most  of these examples are well-known \cite{Kon4, Wang}, the last three equations  are apparently new. We refer to \cite{FMN} for  technical details, Lax pairs, etc. Based on the same method, in \cite{NF} the above classification was extended to  fifth order scalar equations, while
the paper \cite{HN} provides a complete list of integrable two-component systems of  Davey-Stewartson type (all with the same nonlocality $w_x=u_y$). See also \cite{Adler, Mik, ShaYam} and references therein for the existing classification results of integrable systems in 2+1 D. 

\noindent {\bf Remark.} The search for integrable equations (\ref{nonsym}) with `nested' non-localities of the form
$w_x=u_y, \ v_x=f(u, w)_y$ leads to  commuting flows of equations from Theorem  1, and does not give essentially new examples.  

The aim of this work is to extend the above classification to various differential-difference non-localities $w$  such as
$$
\triangle_x^+w=\frac{T_x+1}{2}u_y, ~~~~~  w_x=\triangle_y^+u, ~~~~~   \triangle_x^+w=\triangle_y^+u.
$$
The first two cases are known as  the intermediate long wave and the Toda type non-localities, respectively. We use the following standard  notation for the $\epsilon$-shift operators and the forward/backward discrete derivatives,
$$
T_xf(x, y)=f(x+\epsilon, y), ~~~ T^{-1}_xf(x, y)=f(x-\epsilon, y),
~~~ \triangle_x^+=\frac{T_x-1}{\epsilon}, ~~~ \triangle_x^-=\frac{1-T^{-1}_x}{\epsilon},
$$
 same for $T_y, \  T_y^{-1}, \  \triangle_y^+, \  \triangle_y^-.$ Note that all our non-localities reduce to $w_x=u_y$ in the dispersionless limit $\epsilon \to 0$. 
 
 Here is a brief summary of our classification results (see Sect. 3 for the corresponding Lax pairs, dispersionless limits,  etc).
In Sect. 3.1 we consider two classes of equations with the nonlocality of  intermediate long wave type. Our results are summarised in Theorems \ref{ILW_1} and \ref{ILW_2} below:

\medskip

\begin{theorem}  \label{ILW_1} The following  examples constitute a complete list of integrable equations  of the  form 
$$
u_t=\varphi u_x+\psi u_y +\tau w_x+\eta w_y+\epsilon(...)+\epsilon^2(...),
$$ 
where $w$ is the non-locality of  intermediate long wave type, $\triangle_x^+w=\frac{T_x+1}{2}u_y$, the coefficients $\varphi,\psi,\tau,\eta$ are functions of $u,w$, and
the terms at  $\epsilon$ and $\epsilon^2$ are assumed to be
homogeneous differential polynomials of  the order two and  three
in the $x-$ and $y-$derivatives of $u$ and $w$:
\begin{equation*}
\begin{aligned}
&u_t=uu_y+w_y,\label{ILW1}\\
& u_t=(w+\alpha e^u)u_y+w_y, \\
&u_t=u^2u_y+(uw)_y+\frac{\epsilon^2}{12}u_{yyy}, \\
&u_t=u^2u_y+(uw)_y+\frac{\epsilon^2}{12}\left(u_{yy}-\frac{3}{4}\frac{u_y^2}{u}\right)_y. 
\end{aligned}
\end{equation*}
\end{theorem}

\noindent The first example appeared in \cite{Date} as a differential-difference analogue of the KP equation, see also \cite{Tamizhmani}. It can  be viewed as a $2+1$ dimensional integrable version of the intermediate long wave equation \cite{Zakharov1}.  The  last two examples are differential-difference versions of the Veselov-Novikov and the modified Veselov-Novikov equations, respectively (the third example appeared previously in \cite{Qian}). The second and the fourth equations seem to be new.

\medskip

\begin{theorem}   \label{ILW_2} The following  examples constitute a complete list of integrable equations  of the  form 
$$u_t=\psi u_y+\eta w_y +f\triangle_x^+g+p\triangle_x^-q,$$ 
where $w$ is the non-locality of  intermediate long wave type, $\triangle_x^+w=\frac{T_x+1}{2}u_y$, and
$\psi, \eta, f, g, p, q$ are functions  of $u$ and $w$:
\begin{equation*}
\begin{aligned}
&u_t=uu_y+w_y,\label{ILW1}\\
& u_t=(w+\alpha e^u)u_y+w_y, \\
&u_t= wu_y+w_y+\frac{\triangle_x^++\triangle_x^-}{2} e^{2u}, \\
&u_t= wu_y+w_y+e^u(\triangle_x^++\triangle_x^-) e^{u}. \\
\end{aligned}
\end{equation*}
\end{theorem}

\noindent Here the first two examples are the same as in Theorem 2, while the third 
appeared in \cite{Lom}. The fourth equation is apparently new. 

\bigskip

In Sect. 3.2 we consider the case of integrable equations with the Toda non-locality. The main result is as follows:

\medskip

\begin{theorem}  \label{Toda}  The following  examples constitute a complete list of integrable equations  of the  form 
$$
u_t=\varphi  u_x+f \triangle_y^+g+p \triangle_y^- q,
$$ 
where $w$ is the  non-locality of Toda type, $w_x=\triangle_y^+u$, and $\varphi, f, g, p, q$ are functions  of $u$ and $w$:
\begin{equation*}
\begin{aligned}
& u_t= u\triangle_y^-w,   \\
& u_t=(\alpha u+ \beta) \triangle_y^-e^w,    \\
& u_t= e^w \sqrt{u} \triangle_y^+\sqrt{u}+\sqrt{u}\triangle_y^-(e^w \sqrt{u}),  
\end{aligned}
\end{equation*}
here $\alpha, \beta=const.$
\end{theorem}

\noindent The first example is the well-known Toda equation, while the second  is equivalent to the Volterra (Toda)  chain when $\alpha \ne 0$ ($\alpha=0$), respectively. The third equation seems to be new.

\bigskip

In Sect. 3.3 we discuss the case of integrable equations with two discrete variables. Our main result  is 

\medskip

\begin{theorem}  \label{Discr} The following examples constitute a complete list of integrable equations of the form 
 $$
u_t=f \triangle_x^+ g+h \triangle_x^- k+p\triangle_y^+ q+r \triangle_y^- s,
$$
where  $w$ is the fully discrete non-locality of the form $\triangle_x^+ w=\triangle_y^+ u$, and $f, g, h, k, p, q, r, s$ are functions of $u$ and $w$:
\begin{equation*}
\begin{aligned}
& u_t=u \triangle_y^- (u-w), \\
& u_t=u (\triangle_x^++\triangle_y^-) w, \\
& u_t=(\alpha e^{-u}+\beta) \triangle_y^- e^{u-w}, \\
& u_t=(\alpha e^u+\beta) ( \triangle_x^+ + \triangle_y^-) e^w, \\
& u_t=\sqrt {\alpha- \beta e^{2 u}} \left(e^{w-u} \triangle_y^+ \sqrt {\alpha-\beta e^{2 u}}+ \triangle_y^- (e^{w-u} \sqrt {\alpha-\beta e^{2 u}}) \right),
\end{aligned}
\end{equation*}
here $\alpha, \beta=const$.
\end{theorem}

\noindent In equivalent form, the last example is known as the $2+1$ dimensional analogue of the modified Volterra lattice \cite{Tsuchida}.

Although fully discrete 3D equations are outside the scope of this paper, they can be dealt with in a similar way, suggesting  an alternative to the standard approach based on  the multidimensional consistency   \cite{ Adler1, ABS,  Nijhoff, TW1}. The only constraint required by  our method is that of the non-degeneracy of the  dispersionless limit (see Sect. 2).  Numerous examples of this type are provided by various equivalent forms of the Hirota  equation \cite{Hirota, Miwa} governing KP/BKP/Toda hierarchies. These include the following difference equations (we prefer to write them using discrete derivatives $\triangle$ rather than shifts $T$: in this form  the dispersionless limit becomes more easily seen):
$$
\begin{array}{c}
(\triangle_x^+u-\triangle_y^+u)\triangle_{x}^+\triangle_{y}^+u+
(\triangle_t^+u-\triangle_x^+u)\triangle_{x}^+\triangle_{t}^+u+
(\triangle_y^+u-\triangle_t^+u)\triangle_{y}^+\triangle_{t}^+u=0,\\
\ \\
\triangle^+_x\left(\ln \frac{\triangle^+_tu}{\triangle^+_yu}\right)+\triangle^+_y\left(\ln \frac{\triangle^+_xu}{\triangle^+_tu}\right)+\triangle^+_t\left(\ln \frac{\triangle^+_yu}{\triangle^+_xu}\right)=0,\\
\ \\
\alpha \ e^{\triangle_{x}^+\triangle_{y}^+u}+\beta \ e^{\triangle_{x}^+\triangle_{t}^+u}+\gamma \ e^{\triangle_{y}^+\triangle_{t}^+u}=0, \\
\ \\
\alpha \ e^{\triangle_{x}^+\triangle_{x}^-u}+\beta \ e^{\triangle_{y}^+\triangle_{y}^-u}+\gamma \ e^{\triangle_{t}^+\triangle_{t}^-u}=0, \\
\ \\
\end{array}
$$
etc, here $\alpha, \beta, \gamma$ are arbitrary constants. The corresponding dispersionless limits  $\epsilon \to 0$ can be obtained by replacing discrete derivatives by partial derivatives. This gives
$$
\begin{array}{c}
(u_x-u_y)u_{xy}+(u_t-u_x)u_{xt}+(u_y-u_t)u_{yt}=0,\\
\ \\
\left(\ln \frac{u_t}{u_y}\right)_x+\left(\ln \frac{u_x}{u_t}\right)_y+\left(\ln \frac{u_y}{u_x}\right)_t=0, \\
\ \\
\alpha \ e^{u_{xy}}+\beta \ e^{u_{xt}}+\gamma \ e^{u_{yt}}=0, \\
\ \\
\alpha \ e^{u_{xx}}+\beta \ e^{u_{yy}}+\gamma \ e^{u_{tt}}=0, \\
\end{array}
$$
respectively. In different context, these and similar dispersionless equations appeared in \cite{Bogdanov1, Bogdanov2, BFT, FHK}, see also references therein. One can show that all of the above difference equations inherit hydrodynamic reductions of their dispersionless limits, at least to the order $\epsilon^4$. This property can be used to classify  integrable difference equations within particularly interesting subclasses. As a simple illustration let us consider equations of the form
$$
\triangle_x^+\triangle_x^-(u)+\triangle_y^+\triangle_y^-(f(u))+\triangle_t^+\triangle_t^-(g(u))=0,
$$
where $f$ and $g$ are functions to be determined. The requirement that hydrodynamic reductions of its dispersionless limit, $u_{xx}+f(u)_{yy}+g(u)_{tt}=0$, can be deformed into reductions of the difference equation
up to the order $\epsilon^2$, leads to the following constraints for $f$ and $g$:
$$
f''+g''=0, ~~~ g''(1+f')-g'f''=0, ~~~ f''^2(1+2f')-f'(f'+1)f'''=0.
$$
Without any loss of generality this gives $f(u)=\ln (1-e^u)-u, ~ g(u)=-\ln (1-e^u)$, resulting in the difference equation
$$
\triangle_x^+\triangle_x^-(u)+\triangle_y^+\triangle_y^-(\ln (1-e^u)-u)-\triangle_t^+\triangle_t^-(\ln (1-e^u))=0,
$$
which is yet another equivalent form of the Hirota equation, known as the `gauge-invariant form' \cite{Zabrodin}. Its dispersionless limit appeared recently   in the classification of integrable equations possessing the `central quadric ansatz' \cite{FHZ}. We hope to report further classification results elsewhere.

\section{Classification scheme}

Let us illustrate our approach using the example of the 2+1 dimensional Toda  equation,  
$$
u_t=u\triangle_y^-w, ~~~ w_x=\triangle_y^+u.
$$
Expanding the right hand sides using Taylor's formula  one obtains
\begin{equation}
\begin{aligned}
\label{TL_exp}
\frac{u_{t}}{u} &= w_{y} - \frac{\epsilon}{2} w_{yy} +
\frac{\epsilon^{2}}{6} w_{yyy} +  \dots, \\~~~
w_{x} &= u_{y} + \frac{\ep}{2} u_{yy} + \frac{\ep^{2}}{6} u_{yyy}+ \dots.
\end{aligned}
\end{equation}
The corresponding dispersionless limit results upon setting $\epsilon=0$:
\begin{equation}
\label{dTL}
u_{t} = u w_{y}, \qquad w_{x} = u_{y}.
\end{equation}
This dispersionless system admits exact solutions of the form 
\begin{equation}
\label{uw}
u = R, \qquad w = w(R),
\end{equation}
where $R(x, y, t)$ satisfies the pair of Hopf-type equations,
\begin{equation}
\label{Red} R_y = \mu R_x, \qquad R_{t} =
\mu^2 R R_{x}.
\end{equation}
Here $\mu(R)$ is an arbitrary function, and $w'=\mu$. Solutions of this type are known as one-phase solutions (or planar simple waves, or one-component hydrodynamic reductions). One can show that both relations (\ref{uw}), (\ref{Red}) can be deformed into solutions of the full Toda equation in the form
\begin{equation}
\label{SolutionTL_Def}
u = R, \qquad
w = w(R)
 + \epsilon w_1 R_{x}
+ \epsilon^2 (w_2 R_{xx} + w_3 R_{x}^{2}) + O(\epsilon^{3}),
\end{equation}
and
\begin{gather}
\label{RiemannTL_Def}
\begin{aligned}
R_{y} =& \mu R_{x} +\epsilon^{2} (\alpha_{1} R_{xxx} + \alpha_{2}
R_{x} R_{xx}  +
\alpha_{3} R_{x}^{3}) + O(\epsilon^{4}),\\
R_{t} =&\mu^2 R R_{x}+ \epsilon^{2}(\beta_{1} R_{xxx} +
\beta_{2} R_{x} R_{xx} + \beta_{3} R_{x}^3 ) +
O(\epsilon^{4}),
\end{aligned}
\end{gather}
where $w_i, \alpha_i,  \beta_{i}$ are certain functions of $R$. 
We point out that, modulo the Miura group \cite{Dub1}, the relation $u=R$ can be assumed undeformed. Furthermore, one can show that  odd order  $\epsilon$-corrections  in the equations (\ref{RiemannTL_Def}) (but not (\ref{SolutionTL_Def})) must
vanish identically. Substituting (\ref{SolutionTL_Def})
into~(\ref{TL_exp}), using (\ref{RiemannTL_Def}) and the
compatibility condition $R_{yt} = R_{ty}$,  one can explicitly calculate all coefficients in (\ref{SolutionTL_Def}) and (\ref{RiemannTL_Def}) in terms of
$\mu$ and its derivatives by collecting terms at  different powers of $\epsilon$ \cite{FerM}. This gives
\begin{gather*}
\begin{aligned}
w_1 =& \frac{1}{2}\mu^2, \\
w_2 =&\frac{1}{12} \mu ^2 \left(2 \mu +R \mu '\right), \\
w_3 =&\frac{1}{24} \left(R \left(\mu '\right)^2 \left(2 \mu -R \mu '\right)+\mu ^2 \left(11 \mu '+R \mu ''\right)\right),\\
\alpha_1 =& \frac{1}{12} R \mu ^2 \mu ', \\
\alpha_2 =&\frac{1}{12} R \left(\left(\mu '\right)^2 \left(4 \mu -R \mu '\right)+2 \mu ^2 \mu ''\right),\\
\alpha_3 =& \frac{1}{24} R \left(3 \mu ' \mu '' \left(2 \mu -R \mu '\right)+\mu ^2 \mu ^{(3)}\right),\\
\beta_1 =&\frac{1}{12} R \mu ^3 \left(\mu +2 R \mu '\right), \\
\beta_2 =&\frac{1}{12} R \mu  \left(R \left(\mu '\right)^2 \left(11 \mu -2 R \mu '\right)+4 \mu ^2 \left(3 \mu '+R \mu ''\right)\right),\\
\beta_3 =&\frac{1}{12} R \left(R \left(\mu '\right)^3 \left(2 \mu -R \mu '\right)+8 R \mu ^2 \mu ' \mu ''+\mu  \left(\mu '\right)^2 \left(11 \mu -3 R^2 \mu ''\right)+\mu ^3 \left(4 \mu ''+R \mu ^{(3)}\right)\right),
\end{aligned}
\end{gather*}
etc. We point out that this calculation is an entirely algebraic procedure. Thus, all one-phase solutions of the dispersionless system are `inherited' by the original dispersive equation, at least to the order $O(\ep^{4})$: it is still an open problem to prove their inheritance to all orders in the deformation parameter $\epsilon$. It is important that this works for {\it arbitrary} $\mu(R)$. The requirement of the inheritance of all hydrodynamic reductions of the dispersionless limit by the full dispersive equation is very restrictive (even to  the order $O(\ep^{2})$), and can be used as an efficient classification criterion in the search for integrable equations. In all examples considered so far, the existence of such deformations to the order $\epsilon^4$ was already sufficient for integrability (in many cases, even the order $\epsilon^2$ was enough), and implied the existence of conventional Lax pairs.

\medskip

As an illustration, let us consider the class of Toda-type equations of the form
$$
u_t=f\triangle_y^-g, ~~~ w_x=\triangle_y^+u
$$
where $f(u, w)$ and $g(u, w)$ are two arbitrary functions. Imposing the requirement that all one-phase solutions of the corresponding dispersionless limit  are inherited by the full dispersive equation, we obtain very strong constraints for $f$ and $g$.
Indeed at the order $\epsilon$ we get
$$
 g_u= 0, ~~~
 f_u f_w= 0,  ~~~
 f_w \left(f g_{ww}+g_w f_w\right)=0,
$$
so that, excluding the  degenerate case $f_u=0$, we arrive at
$
g_u=0, \ f_w=0.
$
At the order $\epsilon^2$ one obtains two additional constraints:
$$
 f''(u)= 0,  ~~~
 g''(w)^2-g'(w) g^{'''}(w)=0.
$$
Modulo elementary changes of variables this leads to the cases $f(u)=u, \ g(w)=w$ and 
$f(u)=\alpha u+\beta, \  g(w)=e^w$ which correspond to the Toda and Volterra chains, respectively (see Section 3.2).
Note that all these constraints appear at the order $\epsilon^2$, and are already sufficient for the integrability, implying the existence of  Lax pairs.

\bigskip

\noindent{\bf Remark.} All equations discussed in this paper possess dispersionless limits of the form
\begin{equation}
u_t=\varphi u_x+\psi u_y +\eta w_y, ~~~
 w_x=u_y,
\label{D}
\end{equation}
which will be assumed {\it  non-degenerate} in the following sense:

\noindent (i) The coefficient $\eta$ is nonzero: this is equivalent to the requirement that
the corresponding dispersion relation  defines an
irreducible conic. 

\noindent (ii) The dispersionless limit (\ref{D}) is not  {\it totally linearly
degenerate}. Recall that totally linearly degenerate systems are characterised by the relations \cite{FMN}
$$
\eta_w=0, ~~~ \psi_w+\eta_u=0, ~~~ \varphi_w+\psi_u=0, ~~~
\varphi_u=0.
$$
Dispersive  deformations of degenerate systems  do not inherit  hydrodynamic
reductions, and  require a different
approach which is beyond the scope of this paper. We point out that most of the integrable examples of interest are non-degenerate in the above sense, or can be brought into a non-degenerate form.

\section{Differential-difference equations}

In this section we search for integrable examples within  various classes of differential-difference PDEs generalising intermediate long wave and Toda type equations. We skip the details of calculations, which essentially follow the  pattern outlined in Sect. 2. Classification results are presented modulo Galilean transformations, and transformations of the form $u\to \alpha u+\beta, \ w\to \alpha w+\gamma$.

\subsection{Equations with  the intermediate long wave non-locality: $\triangle_x^+w=\frac{T_x+1}{2}u_y$}

First we classify integrable equations  of the form 
\begin{equation}
u_t=\varphi u_x+\psi u_y +\tau w_x+\eta w_y+\epsilon(...)+\epsilon^2(...),
\label{ILWtype1}
\end{equation}
where $w$ is the non-locality of the intermediate long wave type, $\triangle_x^+w=\frac{T_x+1}{2}u_y$, or, equivalently, $w=\frac{\epsilon}{2}\frac{T_x+1}{T_x-1}u_y$. Here dots denote terms which are homogeneous polynomials of degree two and three in the $x$- and $y$-derivatives of $u$ and $w$, whose coefficients are allowed to be functions of $u$ and $w$.   One can show that all $\epsilon$-terms, as well as all terms containing derivatives with respect to $x$, in particular $\varphi$ and $\tau$, must vanish identically.

\medskip

\noindent {\bf Theorem 2} {\it The following  examples constitute a complete list of integrable equations  of the  form (\ref{ILWtype1}) with the  non-locality of  intermediate long wave type:
\begin{eqnarray}
u_t&=&uu_y+w_y,  \label{ILW1}\\ 
u_t&=&(w+\alpha e^u)u_y+w_y,  \label{ILW2} \\
u_t&=&u^2u_y+(uw)_y+\frac{\epsilon^2}{12}u_{yyy},   \label{ILW3} \\
u_t&=&u^2u_y+(uw)_y+\frac{\epsilon^2}{12}\left(u_{yy}-\frac{3}{4}\frac{u_y^2}{u}\right)_y.  \label{ILW4}
\end{eqnarray}
}

\noindent Although the first two equations are of the first order, they should be viewed as dispersive: the dispersion is contained in the equation for nonlocality. The equation (\ref{ILW1}), which can be written in the form
$$
u_t=uu_y+\frac{\epsilon}{2}\frac{T_x+1}{T_x-1}u_{yy},
$$
first appeared in \cite{Date} as a differential-difference analogue of the KP equation, see also \cite{Tamizhmani}
(we point out that its dispersionless limit does not coincide with that of KP). It can also be viewed as a $2+1$ dimensional integrable version of the intermediate long wave equation \cite{Zakharov1}.  The  equation (\ref{ILW3}) is a differential-difference version of the Veselov-Novikov equation discussed in \cite{Qian}. The last example can be viewed as a differential-difference version of the modified Veselov-Novikov equation. To the best of our knowledge,  equations (\ref{ILW2}) and (\ref{ILW4}) are new.

 Lax pairs, dispersionless limits and dispersionless Lax pairs for the equations from Theorem \ref{ILW_1} are  provided in the table below (note that  equations (\ref{ILW3}) and (\ref{ILW4}) have coinciding dispersionless limits/dispersionless Lax pairs). Here and in what follows, Lax pairs were obtained by the quantisation of dispersionless Lax pairs as discussed in  \cite{Zakharov}.
\begin{center}
\begin{tabular}{ | l | l | l | p{3cm} |} \hline
 $Eqn$ & $Lax~pair$ & $Dispersionless$ & $Dispersionless$ \\ 
 $$ &  & $limit$ & $Lax~pair$  \\ \hline 
 &  &  & \\
(\ref{ILW1}) & $T_x\psi=\epsilon \psi_y-u\psi$ & $u_t=u u_y+w_y$ & $e^{S_x}=S_y-u$ \\
 & $\epsilon \psi_t=\frac{\epsilon^2}{2}\psi_{yy}+(w-\frac{\epsilon}{2}u_y)\psi$ & $w_x=u_y$ & $S_t=\frac{1}{2}S_y^2+w$ \\ 
 &  &  & \\ \hline
 &  &  & \\
(\ref{ILW2}) & $T_x\psi=\epsilon e^{-u}\psi_y-\alpha \psi  $ & $u_t=(w+\alpha e^u)u_y+w_y$ & $e^{S_x}=e^{-u}S_y-\alpha$ \\
 & $\psi_t=\frac{\epsilon}{2}\psi_{yy}+(w-\frac{\epsilon}{2}u_y)\psi_y$ & $w_x=u_y$ & $S_t=\frac{1}{2}S_y^2+wS_y$ \\ 
 &  &  & \\ \hline
  &  &  & \\
(\ref{ILW3}) & $\epsilon(T_x-1)\psi_y=-2u(T_x+1)\psi$ & $u_t=u^2u_y+(uw)_y$ & $\frac{e^{S_x}-1}{e^{S_x}+1}S_y=-2u$ \\
 & $\psi_t=\frac{\epsilon^2}{12}\psi_{yyy}+(w-\frac{\epsilon}{2}u_y)\psi_y$ & $w_x=u_y$ & $S_t=\frac{1}{12}S_y^3+wS_y$ \\ 
 &  &  & \\ \hline
&  &  & \\
(\ref{ILW4}) & $\epsilon(T_x-1)\psi_y=\frac{\epsilon}{2}\frac{u_y}{u}(T_x-1)\psi-2u(T_x+1)\psi $ & $u_t=u^2u_y+(uw)_y$ & $\frac{e^{S_x}-1}{e^{S_x}+1}S_y=-2u $ \\
& $\psi_t=\frac{\epsilon^2}{12}\psi_{yyy}+(w-\frac{\epsilon}{2}u_y)\psi_y+\frac{1}{2}(w_y-\frac{\epsilon}{2}u_{yy})\psi$  & $w_x=u_y$ & $S_t=\frac{1}{12}S_y^3+wS_y$ \\ 
 &  $$ &  &  \\ \hline
\end{tabular}
\end{center}

\bigskip

\noindent{\bf Remark. } Equations (\ref{ILW1}) and (\ref{ILW2}) are related by a (rather non-trivial) gauge transformation. Let us begin with the dispersionless limit of (\ref{ILW2}), 
$$
u_t=(w+\alpha e^u)u_y+w_y, ~~~ w_x=u_y,
$$
with the corresponding Lax pair 
$$
S_t=\frac{1}{2}S_y^2+wS_y, ~~~ e^{S_x}=e^{-u}S_y-\alpha.
$$
Let $h$ be a potential such that $u=h_x, \ w=h_y$. One can verify that the new variables $\tilde u=w+\alpha e^u, \ \tilde w=h_t-\frac{w^2}{2}, \ \tilde S=S+h$ satisfy the dispersionless equation (\ref{ILW1}),
$$
\tilde u_t=\tilde u\tilde u_y+\tilde w_y, ~~~ \tilde w_x=\tilde u_y,
$$ 
along with the corresponding Lax pair
$$
\tilde S_t=\frac{1}{2}\tilde S_y^2+\tilde w, ~~~ e^{\tilde S_x}=\tilde S_y-\tilde u,
$$
thus establishing the required link at the dispersionless level (it is sufficient to perform this calculation at the level of Lax pairs: the equations for $\tilde u, \tilde w$ will be automatic). The dispersive version of this construction is as follows. We take the equation (\ref{ILW2}),
$$
u_t=(w+\alpha e^u)u_y+w_y, ~~~ w=\frac{\epsilon}{2}\frac{T_x+1}{T_x-1}u_y,
$$
with the corresponding Lax pair
$$
\psi_t=\frac{\epsilon}{2}\psi_{yy}+(w-\frac{\epsilon}{2}u_y)\psi_y, ~~~ T_x\psi=\epsilon e^{-u}\psi_y-\alpha \psi.
$$
Let $H$ be a potential such that $u=\frac{T_x-1}{\epsilon}H, \ w=\frac{T_x+1}{2}H_y$. One can verify that the new variables $\tilde u=H_y+\alpha e^u, \ \tilde w=H_t-\frac{H_y^2}{2}+\frac{\alpha \epsilon}{2} e^u\triangle_x^+H_y, \ \tilde \psi=e^{H/\epsilon}\psi$ satisfy the equation (\ref{ILW1}),
$$
\tilde u_t=\tilde u\tilde u_y+\tilde w_y, ~~~ \tilde w=\frac{\epsilon}{2}\frac{T_x+1}{T_x-1}\tilde u_y,
$$
with the corresponding Lax pair
$$
\epsilon \tilde \psi_t=\frac{\epsilon^2}{2}\tilde \psi_{yy}+(\tilde w-\frac{\epsilon}{2}\tilde u_y)\tilde \psi, ~~~ T_x\tilde \psi=\epsilon \tilde \psi_y-\tilde u\tilde \psi.
$$
Again, it is sufficient to perform this calculation at the level of Lax pairs. Due to the complexity of this transformation we prefer to keep both equations in the list of Theorem \ref{ILW_1} as separate cases. 

\bigskip

\bigskip

Another interesting class of  equations with the non-locality of  intermediate long wave type is
\begin{equation}
u_t=\psi u_y +\eta w_y+f\triangle_x^+g+p\triangle_x^-q,
\label{ILWtype2}
\end{equation}
where  $\triangle_x^+w=\frac{T_x+1}{2}u_y$, and
$\psi, \eta, f, g, p, q$ are functions  of $u$ and $w$.

\medskip

\noindent{\bf Theorem 3}  {\it The following  examples constitute a complete list of integrable equations  of the  form (\ref{ILWtype2}) with the  non-locality of  intermediate long wave type:
\begin{eqnarray}
u_t&=&uu_y+w_y, \nonumber \\ 
u_t&=&(w+\alpha e^u)u_y+w_y, \nonumber \\
u_t&=& wu_y+w_y+\frac{\triangle_x^++\triangle_x^-}{2} e^{2u},   \label{ILW23} \\
u_t&=&wu_y+w_y+e^u(\triangle_x^++\triangle_x^-) e^{u}.  \label{ILW24}
\end{eqnarray}
}

\noindent Here the first two equations are the same as in Theorem \ref{ILW_1}, the third example first appeared in \cite{Lom}, while the fourth is apparently new.  Lax pairs, dispersionless limits and dispersionless Lax pairs for equations from Theorem \ref{ILW_2} are  provided in the table below (note that equations (\ref{ILW23}) and (\ref{ILW24}) have coinciding dispersionless limits):
\begin{center}
\begin{tabular}{ | l | l | l | p{3.5cm} |} \hline
 $Eqn$ & $Lax~pair$ & $Dispersionless$ & $Dispersionless$ \\ 
 $$ &  & $limit$ & $Lax~pair$  \\ \hline 
 &  &  & \\
(\ref{ILW23}) & $\epsilon \psi_y=(T_xe^{u})T_x\psi+e^uT_x^{-1}\psi $ & $u_t=2e^{2u}u_x+wu_y+w_y$ & $S_y=2e^u\cosh S_x$ \\
 & $\epsilon \psi_t=\frac{1}{2}e^{T_x(1+T_x)u}T_x^2\psi-\frac{1}{2}e^{(1+T_x^{-1})u}T_x^{-2}\psi+      $ & $w_x=u_y$ & $S_t=e^{2u}\sinh 2S_x +$ \\ 
 & $~~~~~~~~T_x(we^{u})T_x\psi+we^uT_x^{-1}\psi$ & $$ & $~~~~~~~2we^u\cosh S_x$\\
 &  &  & \\ \hline
&  &  & \\
(\ref{ILW24}) & $\epsilon \psi_y=e^u (T_x\psi+T_x^{-1}\psi) $ & $u_t=2e^{2u}u_x+wu_y+w_y$ & $S_y=2e^u\cosh S_x$ \\
 & $\epsilon \psi_t=\frac{1}{2}e^{(1+T_x)u}T_x^2\psi-\frac{1}{2}e^{(1+T_x^{-1})u}T_x^{-2}\psi+      $ & $w_x=u_y$ & $S_t=e^{2u}\sinh 2S_x+$ \\ 
 & $w e^u(T_x\psi+T_x^{-1}\psi)+\frac{\epsilon}{2} e^u [(\triangle_x^++\triangle_x^-)e^u]\psi$ & $$ & $~~~~~~~ 2we^u\cosh S_x$\\
 &  &  & \\ \hline
\end{tabular}
\end{center}

\bigskip

\subsection{ Equations with the Toda non-locality: $w_x=\triangle_y^+u$ }

In this section we  classify integrable equations of the form
\begin{equation}
u_t=\varphi u_x+f \triangle_y^+g+p \triangle_y^- q,
\label{toda}
\end{equation}
where the non-locality $w$ is defined as  $w_x=\triangle_y^+u$, and $\varphi, f, g, p, q$ are functions of $u$ and $w$. 
 
\medskip
 
\noindent{\bf Theorem 4} {\it The following  examples constitute a complete list of integrable equations  of the  form (\ref{toda}) with the  non-locality of Toda type:
\begin{eqnarray}
u_t&=& u\triangle_y^-w,   \label{Toda1} \\
u_t&=& (\alpha u+ \beta) \triangle_y^-e^w,   \label{Toda2} \\
u_t&=& e^w \sqrt{u} \triangle_y^+\sqrt{u}+\sqrt{u}\triangle_y^-(e^w \sqrt{u}),   \label{Toda3} 
\end{eqnarray}
here $\alpha, \beta=const$.}

\medskip

 Equation (\ref{Toda1})  is the 2+1 dimensional Toda equation, which can also be written in the form $(\ln u)_{xt}=\triangle_y^+ \triangle_y^- u$, while equation (\ref{Toda2}) is equivalent  to the Volterra chain when $\alpha \ne 0$, or to the Toda chain when $\alpha=0$.
Lax pairs, dispersionless limits and dispersionless Lax pairs for the equations from Theorem \ref{Toda} are  provided in the table below:
\begin{center}
\begin{tabular}{ | l | l | l | p{4cm} |} \hline
 $Eqn$ & $Lax~pair$ & $Dispersionless$ & $Dispersionless$ \\ 
 $$ &  & $limit$ & $Lax~pair$  \\ \hline 
 &  &  & \\
(\ref{Toda1}) & $\epsilon T_y\psi_x=u\psi$ & $u_t=uw_y$ & $e^{S_y}S_x=u$ \\
 & $\epsilon\psi_t=-T_y\psi+(T_y^{-1}w)\psi $ & $w_x=u_y$ & $S_t=-e^{S_y}+w$ \\ 
 &  &  & \\ \hline
&  &  & \\
(\ref{Toda2}) & $\epsilon T_y\psi_x=-(T_y u) T_y\psi+(\alpha T_y u+\beta)\psi$ & $u_t=(\alpha u + \beta) e^w w_y$ & $e^{S_y}S_x=-u e^{S_y}+\alpha u +\beta$ \\
 & $\epsilon\psi_t=-e^w T_y\psi +\alpha e^w\psi$ & $w_x=u_y$ & $S_t=-e^w e^{S_y}+\alpha e^w$ \\ 
 &  &  & \\ \hline
&  &  & \\
(\ref{Toda3}) & $\epsilon T_y\psi_x=\epsilon \sqrt{\frac{T_y u}{u}} \psi_x-(T_y u)T_y \psi-\sqrt{u T_y u}~ \psi$ & $u_t=e^w u_y+u e^w w_y$ & $e^{S_y}S_x=S_x-u e^{S_y}-u$ \\
 & $\epsilon \psi_t=\frac{1}{2} e^w T_y \psi-\frac{1}{2} (T_y^{-1}e^w) T_y^{-1} \psi$ & $w_x=u_y$ & $S_t=e^w \sinh S_y$ \\ 
 &  &  & \\ \hline
\end{tabular}
\end{center}

\medskip

\noindent{\bf Remark}. One can show that there exist no non-degenerate integrable equations of the form
$$
u_t=\eta w_y+f \triangle_x^+g+p \triangle_x^- q,
$$
where the non-locality $w$ is defined as  $\triangle_x^+w=u_y$, and $\eta, f, g, p, q$ are functions of $u$ and $w$. Indeed, the integrability requirement implies the condition $\eta=0$, which corresponds to degenerate systems.

\bigskip

\subsection{Equations with the fully discrete non-locality: $\triangle_x^+w=\triangle_y^+u$}

In this last section we classify integrable  equations of the form
\begin{equation}
\label{discr}
\begin{aligned}
& u_t=f \triangle_x^+ g+h \triangle_x^- k+p\triangle_y^+ q+r \triangle_y^- s, \\
\end{aligned}
\end{equation}
where the non-locality $w$ is defined as $ \triangle_x^+ w=\triangle_y^+ u$, and the functions $f, g, h, k, p, q, r, s$ depend on $u$ and $w$.

\medskip

\noindent{\bf Theorem 5}  {\it The following examples constitute a complete list of integrable equations of the form (\ref{discr}) with the fully discrete non-locality:
\begin{eqnarray}
 u_t&=&u \triangle_y^- (u-w), \label{Discr1} \\
 u_t&=&u (\triangle_x^++\triangle_y^-) w, \label{Discr2}\\
 u_t&=&(\alpha e^{-u}+\beta) \triangle_y^- e^{u-w}, \label{Discr4}\\
 u_t&=&(\alpha e^u+\beta) ( \triangle_x^+ + \triangle_y^-) e^w, \label{Discr5}\\
 u_t&=&\sqrt {\alpha- \beta e^{2 u}} \left(e^{w-u} \triangle_y^+ \sqrt {\alpha-\beta e^{2 u}}+ \triangle_y^- (e^{w-u} \sqrt {\alpha-\beta e^{2 u}}) \right),\label{Discr6}
\end{eqnarray}
here $\alpha, \beta=const$.
}

\medskip

\noindent In equivalent form, equation (\ref{Discr6}) is known as the $2+1$ dimensional analogue of the modified Volterra lattice \cite{Tsuchida}.   Lax pairs, dispersionless limits and dispersionless Lax pairs for the equations from Theorem \ref{Discr} are  provided in the table below:
\begin{center}
\begin{tabular}{ | l |  l | l |p{3.6cm} |} \hline
 $Eqn$ & $Lax~pair$ & $Dispersionless$ & $Dispersionless$ \\ 
 $$ &  & $limit$ & $Lax~pair$  \\ \hline 
 &  &  & \\
(\ref{Discr1}) & $T_x T_y \psi=-T_y \psi+(T_y u) T_x\psi$ & $u_t=u (u_y-w_y)$ & $e^{S_x+S_y}=-e^{S_y}+u e^{S_x}$ \\
 & $\epsilon \psi_t=T_y \psi-w\psi$ & $w_x=u_y$ & $S_t=e^{S_y}-w$ \\ 
 &  &  & \\ \hline
 &  &  & \\
(\ref{Discr2}) & $T_xT_y \psi=T_y \psi-u\psi$ & $u_t=u (u_y+w_y)$ & $e^{S_x+S_y}=e^{S_y}-u$ \\
 & $\epsilon \psi_t=T_y \psi+(T_y^{-1}w) \psi$ & $w_x=u_y$ & $S_t=e^{S_y}+w$ \\ 
 &  &  & \\ \hline
&  &  & \\
(\ref{Discr4}) & $T_y^{-1} \psi=\frac{e^u}{\alpha+\beta e^u} T_x^{-1} \psi+\frac{1}{\alpha+\beta e^u} \psi$ & $u_t=(\alpha+\beta e^u)e^{-w}(u_y-w_y)$ & $e^{-S_y}=\frac{e^u e^{-S_x}+1}{\alpha+\beta e^u}$ \\
& $\epsilon T_x^{-1} \psi_t=-\epsilon e^{-u} \psi_t-$ & $w_x=u_y$ & $e^{-S_x} S_t=-e^{-u} S_t-$ \\ 
 & $~~~~\alpha e^{-w} T_x^{-1} \psi+\beta e^{-w} \psi$ &  & $~~\alpha e^{-w} e^{-S_x}+\beta e^{-w}$ \\ 
&  &  & \\ \hline
&  &  & \\
(\ref{Discr5}) & $T_y^{-1} \psi=-\frac{e^u}{\alpha e^u+\beta} T_x \psi+\frac{1}{\alpha e^u+\beta} \psi$ & $u_t=(\alpha e^u+\beta) e^w (u_y+w_y)$ & $e^{-S_y}=\frac{-e^u e^{S_x}+1}{\alpha e^u+\beta}$ \\
& $\epsilon T_x \psi_t=\epsilon e^{-u} \psi_t-\beta (T_x e^{w}) T_x \psi-$ & $w_x=u_y$ & $e^{S_x} S_t=e^{-u} S_t-$ \\ 
 & $~~~~~~~~~~~\alpha (T_x e^{w}) \psi$ &  & $~~~\beta e^{w} e^{S_x}-\alpha e^{w}$ \\ 
&  &  & \\ \hline
&  &  & \\
(\ref{Discr6}) & $T_x T_y \psi=\frac{\alpha}{\beta} (T_y e^{-u})T_y \psi+$ & $u_t=\alpha (e^{w-u})_y-\beta (e^{w+u})_y$ & $e^{S_x+S_y}=\frac{\alpha}{\beta} e^{-u} e^{S_y}+ $ \\
& $~~~~~~~\frac{T_y(e^{-u} \sqrt{\alpha-\beta e^{2u}})}{\sqrt{\alpha-\beta e^{2u}}}\left (T_x \psi- e^u \psi \right)$ & $w_x=u_y$ & $~~~~~~~~~~~~e^{-u} e^{S_x}-1$ \\
& $\epsilon \psi_t=-\alpha e^w T_y \psi+\beta (T_y^{-1}e^w) T_y^{-1} \psi$ &  & $S_t= -\alpha e^w e^{S_y}+$ \\ 
 &  &  & $~~~~~~~~\beta e^we^{-S_y}$ \\ \hline

\end{tabular}
\end{center}

\medskip

\noindent{\bf Remark 1.}  The continuum limit of the modified Volterra lattice (\ref{Discr6}) in $x-$direction, namely  $x\to h x, ~u\to hu$ and $h \to 0$, gives the Toda-type  lattice (\ref{Toda3}). Similarly, in the same limit equations (\ref{Discr1}) and (\ref{Discr2}) give the Toda equation (\ref{Toda1}), while the remaining two, (\ref{Discr4}) and (\ref{Discr5}), lead to  the equation (\ref{Toda2}) with $\alpha=0$.

\medskip

\noindent{\bf Remark 2.} We point out that there exist other types of integrable equations with the  non-locality $\triangle_x^+w=\triangle_y^+u$, which are not covered by Theorem \ref{Discr}. One of such examples  is the first flow of the discrete modified Veselov-Novikov hierarchy constructed in \cite{DZakharov}, 
$$
u_t=\sqrt{-(T_y^{-1}\triangle_x^+e^{2w})(\triangle_x^+e^{-2w})}, ~~~ \triangle_x^+w=\triangle_y^+u.
$$
This equation is not of the form (\ref{discr}), furthermore, its dispersionless limit is degenerate: $$u_t=2w_x, \ w_x=u_y.$$

\section{Concluding remarks}

This paper  outlines an approach to the classification of integrable  differential-difference equations in 2+1D (with one/two discrete variables) based on the method of hydrodynamic reductions.

\medskip

\noindent {\bf 1.} It would be challenging to extend our approach to the classification of  fully discrete 3D equations.  Some work in this direction, based on the concept of multidimensional consistency, has been done in \cite{Adler1,  TW1}.

\medskip

\noindent {\bf 2.} A novel approach to the integrability of multidimensional lattices of the Toda and Volterra type, based on the concept of characteristic Lie rings, was proposed recently in \cite{Habibullin}. Its relation to our method is yet to  be properly investigated. 

\medskip

\noindent {\bf 3.} Our calculations demonstrate that, even though nontrivial combinations of differential and difference operators (in the same independent variable) were allowed in the initial ansatz, non of such `differential-delay'  cases survived the integrability test, leading to either purely discrete, or purely differential situations. It remains to be seen whether there exist proper differential-delay integrable equations in 2+1 D.

\section*{Acknowledgements}

We  thank A. Mikhailov, M. Pavlov and  J.-P. Wang for useful discussions. The research of EVF was partially supported  by the European Research Council Advanced Grant  FroM-PDE.

\end{document}